\documentclass[lettersize,journal]{IEEEtran}
\usepackage{amsmath,amsfonts}
\usepackage{array}
\usepackage{textcomp}
\usepackage{stfloats}
\usepackage{url}
\usepackage{verbatim}
\usepackage{graphicx}
\usepackage{cite}
\usepackage{amssymb}
\usepackage{caption}
\usepackage{tikz}
\usepackage{titlesec}
\usepackage{subcaption}
\usepackage{hyperref}
\usepackage{amsthm}
\usepackage{textcomp}
\usepackage{xcolor}
\usepackage{textcomp}
\usepackage{stfloats}
\usepackage{url}
\usepackage{verbatim}
\usepackage{gensymb}
\usepackage[linesnumbered,ruled,vlined]{algorithm2e}
\def\BibTeX{{\rm B\kern-.05em{\sc i\kern-.025em b}\kern-.08em
    T\kern-.1667em\lower.7ex\hbox{E}\kern-.125emX}}
\usepackage{setspace}
\begin{document}

\title{Inter-RIS Beam Focusing Codebook Design in Cooperative Distributed RIS Systems}

\author{Youssef Hussein, Mohamad Assaad, and Thierry Clessienne

\thanks{Y. Hussein and M. Assaad are with Université Paris-Saclay, CNRS, CentraleSupélec, Laboratoire des signaux et systèmes, Gif-sur-Yvette, France (e-mail:\{youssef.hussein, mohamad.assaad\}@centralesupelec.fr) }

\thanks{Y. Hussein and T. Clessienne are with Orange Labs, Ch\^atillon, France (e-mail:\{youssef.hussein, thierry.clessienne\}@orange.com) }}



\maketitle

\begin{abstract}
This paper explores distributed Reconfigurable Intelligent Surfaces (RISs) by introducing a cooperative dimension that enhances adaptability and performance. It focuses on strategically deploying multiple RISs to improve connectivity with the Base Station (BS) and among RISs, thereby aiding users in areas with weak BS coverage and enhancing spatial multiplexing gains. Each RIS can function as a primary surface to directly support users or as an intermediary surface to reflect signals to another primary surface. This dual functionality enables flexible responses to changing conditions. We implement an inter-RIS signal focusing design for phase shifts, creating a tailored codebook for precise control over signal direction. This design considers the interplay of incidence and reflection angles to maximize reflected signal power, based on the RIS response function and the physical properties of the RIS elements.
\end{abstract}


\section{Introduction}

\IEEEPARstart{T}{he} evolution of wireless communication is marked by continuous innovation to meet the demand for higher data rates, better coverage, and efficient spectrum use. Despite advancements in Fifth Generation (5G), challenges persist, prompting a focus on the development of Sixth Generation (6G) networks. A key technology in this evolution is Reconfigurable Intelligent Surfaces (RISs), which consist of programmable meta-materials that dynamically manipulate signals to enhance communication performance by improving transmission rates and covering signal gaps \cite{55}.

Recent studies on multi-RIS aided Multi-User-MIMO (MU-MIMO) communications have shifted from single-reflection to multi-reflection designs. Research has explored cooperative beamforming for multiple RIS-assisted systems, optimizing user sum rates \cite{118}, and proposed multi-path beam routing schemes to maximize received signal power \cite{144}. In previous work \cite{170}, we addressed low rank RIS-BS channels impacting MU-MISO performance and proposed a distributed RIS system to enhance spatial multiplexing gains. This study focused on optimizing the configuration of RISs to maximize user ergodic sum rates. However, the potential of cooperative RIS frameworks, where RISs can switch roles between primary and intermediary, remains underexplored, limiting adaptability in dynamic environments.

This paper presents a cooperative RIS framework that allows each RIS to function as either a primary or intermediary surface. This dual functionality enhances spatial multiplexing gains and coverage under varying conditions. To reduce the costs of reconfiguring all RISs in different modes, especially in dense deployments, we focus on designing the inter-RIS focusing codebook. This codebook contains codewords for phase shift configurations that focus signals towards specific RISs, using both linear and optimization-based phase shifts to maximize reflected signal energy and enhance the cooperative capabilities of the RIS network.

\textit{Notation:} Vectors and matrices are denoted by boldface lower case and upper case letters, respectively. The conjugate transpose and transpose of a matrix $\mathbf{X}$ are denoted by $\mathbf{X}^H$ and $\mathbf{X}^T$, respectively. $\mathbf{I}_N$ denotes the $N\times N$ identity matrix and $tr(\mathbf{X})$ the trace of a matrix $\mathbf{X}$.

\section{System Model} \label{SM&PF}

We consider a single-cell downlink MU-MIMO system enhanced by $I$ distributed cooperative RISs to extend coverage to areas with poor or no direct BS coverage. A BS with $M$ transmit antennas serves $K$ single-antenna users distributed in $I$ low coverage zones, each served with an RIS. Each RIS $i \in \{1, \dots, I\}$ has $N_i$ passive reflecting elements that can dynamically reflect signals towards nearby users or other RISs. RISs can be positioned on building facades, ensuring LoS propagation to the BS and other RISs, allowing single and double reflections to enhance and control spatial multiplexing gains from one low-coverage zone to another as will be discussed in the sequel. Channels are assumed to be flat fading. The channel vector between the $i$-th RIS and the $k$-th user is $\mathbf{h}_{rik} \in \mathbb{C}^{1 \times N_i}$, where $k \in \{1, \dots, K\}$. The channel matrices between the BS and the $i$-th RIS, and between the $j$-th and $i$-th RIS, are $\mathbf{H}_{Bi} \in \mathbb{C}^{N_i \times M}$ and $\mathbf{H}_{ji} \in \mathbb{C}^{N_i \times N_j}$, respectively. Direct BS-user links are assumed to be obstructed.

As the RISs are installed on high-rise buildings, the BS-RIS and inter-RIS channels exhibit LoS propagation, but scatterers can introduce multi-path arrivals, leading to rank-deficient channels \cite{155}. We adopt a low-rank channel model for both $\mathbf{H}_{Bi}$ and $\mathbf{H}_{ji}$ as follows: $ \mathbf{H}_{Bi} = \mathbf{A}_{Bi} \mathbf{\Sigma}_{Bi} \mathbf{D}_{Bi}^H \in \mathbb{C}^{N_i \times M}$, and $  \mathbf{H}_{ij} = \mathbf{A}_{ij} \mathbf{\Sigma}_{ij} \mathbf{D}_{ij}^H \in \mathbb{C}^{N_j \times N_i}$. Here, $L_{Bi}$ and $L_{ij}$ denote the number of independent paths for the BS-to-RIS $i$ and RIS $i$-to-RIS $j$ channels, respectively. Each channel has one dominant LoS path and $L_{Bi} - 1$ and $L_{ij} - 1$ weaker paths due to scattering. $\mathbf{A}_{Bi} \in \mathbb{C}^{N_i \times L_{Bi}}$ and $\mathbf{A}_{ij} \in \mathbb{C}^{N_j \times L_{ij}}$ are the receive steering matrices characterized by the angles of arrival (AoA) for each path. $\mathbf{D}_{Bi} \in \mathbb{C}^{M \times L_{Bi}}$ and $\mathbf{D}_{ij} \in \mathbb{C}^{N_i \times L_{ij}}$  are the transmit steering matrices, characterized by the angles of departure (AoD) for each path.  $\mathbf{\Sigma}_{Bi} \in \mathbb{C}^{L_{Bi} \times L_{Bi}}$ and $\mathbf{\Sigma}_{ij} \in \mathbb{C}^{L_{ij} \times L_{ij}}$ are diagonal matrices containing the complex channel gains. The columns of  $\mathbf{A}_{Bi}$,  $\mathbf{A}_{ij}$,  $\mathbf{D}_{Bi}$, and $\mathbf{D}_{ij}$ are array steering vectors for each path. An array steering vector $\mathbf{a}_X(\mathbf{\Psi})$ for a Uniform Rectangular Planar Array (UPRA) with $X= X_1 \times X_2$ elements is defined as:  
\begin{align}
&\mathbf{a}_X(\mathbf{\Psi}) = \mathbf{a}_{X_1}\big(A_{x}(\mathbf{\Psi})\big)\otimes \mathbf{a}_{X_2}\big(A_{z}(\mathbf{\Psi}))\big),\\
&\mathbf{a}_{X_1}\big(A_{x}(\mathbf{\Psi})\big) =(1 \,\,\, e^{j2A_{x}}\dots\,\,e^{j2(X_1-1)A_{x}})^T,\\
&\mathbf{a}_{X_2}\big(A_{z}(\mathbf{\Psi})\big) =(1 \,\,\, e^{j2A_{z}}\dots\,\,e^{j2(X_2-1)A_{z}})^T,
\end{align} 
with $\mathbf{\Psi} = (\theta, \phi)$ defined as a pair of elevation and azimuth angles, we also define:
\begin{align}\label{Axyz}
    (A_{x}, A_{y}, A_{z}) &= \big(\sin(\theta)\cos(\phi), \sin(\theta)\sin(\phi), \cos(\theta)\big),
\end{align}
for both transmit ($\mathbf{\Psi}_t$) and receive ($\mathbf{\Psi}_r$) angles, and:
\begin{align}
    A_{l}(\mathbf{\Psi}_t, \mathbf{\Psi}_r) &= A_{l}(\mathbf{\Psi}_t) + A_{l}(\mathbf{\Psi}_r), \forall l \in \{x, y, z\}.
\end{align}

\begin{figure}[t] 
\centering 
\includegraphics[width=40mm]{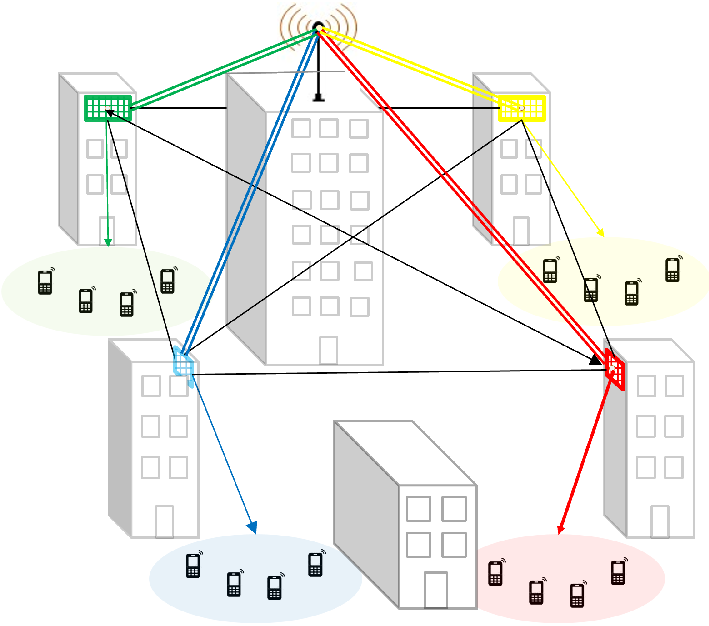}  
\caption{System Model.} 
\label{system2} 
\end{figure}

\begin{figure}[t] 
\centering
\includegraphics[width=40mm]{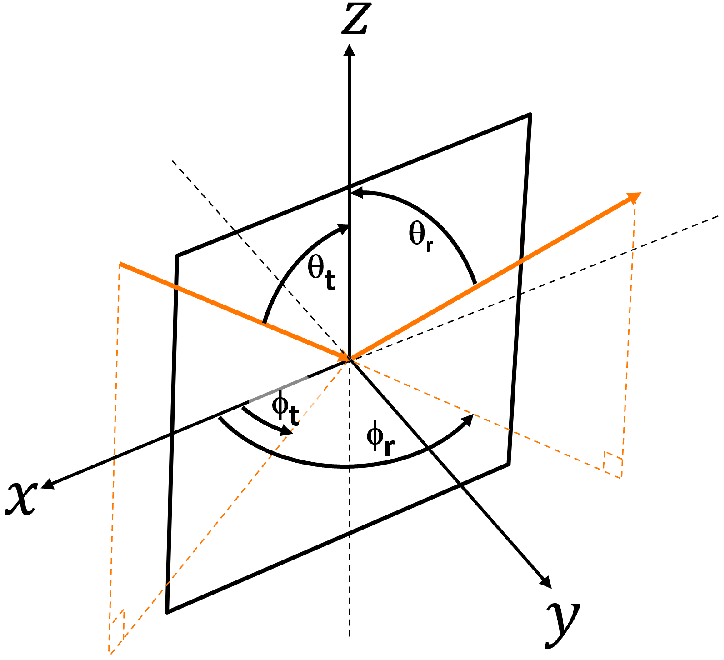} 
\caption{Incidence and reflection angles on RIS.} 
\label{RISfig} 
\end{figure}

Considering the BS, an RIS $i$, and an RIS $j$ such that $i, j \in {1, \dots, I}$, $i \neq j$, we define $\mathbf{\Psi}_{t_{BSi}}^{l_1}$ as the AoD from BS to $i$-th RIS for the $l_1$-th ray, $\mathbf{\Psi}_{r_{BSi}}^{l_1}$ as the AoA at $i$-th RIS from BS for the $l_1$-th ray, $\mathbf{\Psi}_{t_{ij}}^{l_2}$ as the AoD from $i$-th RIS to $j$-th RIS for the $l_2$-th ray, and $\mathbf{\Psi}_{r_{ij}}^{l_2}$ as the AoA at $j$-th RIS from $i$-th RIS for the $l_2$-th ray, where $l_1 \in \{1, \dots, L_{Bi}\}$ and $l_2 \in \{1, \dots, L_{ij}\}$. The RIS is considered a URPA surface with $N_i = N_{ix}N_{iz}$ reflecting elements along the $x$ and $z$ axes. $\mathbf{\Psi}_t = (\theta_t, \phi_t)$ and $\mathbf{\Psi}_r = (\theta_r, \phi_r)$ denote the incident and reflection angles, respectively, as illustrated in Fig. \ref{RISfig}. The phase shift matrix of RIS $i$ is given by: $ \Phi_i=\Bar{g}_{uc_i}\text{diag}(e^{j\varphi_{i,1}},\dots,e^{j\varphi_{{i,N_i}}})$, where $\Bar{g}_{uc_i}$ is the unit-less unit-cell factor characterizing the unit-cell radiation pattern as a function of the physical properties of the unit cell of an RIS $i$ \cite{140}. The $i$-th RIS response function for a signal arriving from AoA $\mathbf{\Psi}_t$ with phase shifts $e^{j\varphi_{{i,n_i}}}$ applied by each of the $n_i\in\{0,...,N_i-1\}$ RIS elements, towards an RIS $j$ with AoD $\mathbf{\Psi}_r$ is:\begin{align} \label{gr}
    g_i(\mathbf{\Psi}_t,\mathbf{\Psi}_r) =& \Bar{g}_{uc_i}\sum_{n_{ix}=0}^{N_{ix}-1}\sum_{n_{iz}=0}^{N_{iz}-1}e^{j\kappa  d_xA_x(\mathbf{\Psi}_t, \mathbf{\Psi}_r)n_{ix}} \nonumber\\
    &\times e^{j\kappa d_zA_z(\mathbf{\Psi}_t,\mathbf{\Psi}_r)n_{iz}} e^{j\varphi_{{i,n_{ix},n_{iz}}}}.
\end{align}
Let $\kappa = \frac{2\pi}{\lambda_c}$, where $\lambda_c$ is the wavelength, and $\Bar{g}_{uc_i} = \frac{4\pi L_{xi}L_{zi}}{\lambda_c^2}$ with $L_{xi}$ and $L_{zi}$ denoting the dimensions of a rectangular RIS unit cell. The inter-element spacing along the $x$ and $z$ directions are denoted by $d_x$ and $d_z$. For simplicity, we assume a constant unit-cell factor $\Bar{g}_{uc_i}$, but it can vary based on the incident and reflected wave angles \cite{140}. The results can be easily adapted to different unit-cell factors as described in \cite{157}.

\section{Cooperative RIS Focusing Codebook Design} \label{codebooksec}
This section details the methodology for configuring RIS phase shifts to optimize signal focusing. The cooperative RISs system can operate within a dynamic framework, selecting a primary surface to reflect signals to a designated area, while other surfaces reflect signals from the BS to the primary RIS. Precise phase shift configuration ensures coherent signal direction, allowing the primary RIS to enhance spatial multiplexing gains and serve more users, improving system capacity and efficiency. This coordination optimizes signal propagation and enhances network robustness and adaptability. However, configuring phase shifts for each intermediary RIS can be computationally intensive, particularly when a large number of surfaces are installed. To address this, we propose a signal focusing codebook for each RIS based on the response function defined in \eqref{gr}, which depends on the physical properties and angles of arrival and departure. Since BSs and RISs are fixed, LoS angles remain constant, while multipath components change slowly. Each RIS's codebook contains $I-1$ codewords, each focusing signals towards one of the other RISs. By optimizing phase shifts to maximize the response function, we ensure efficient signal reflection. This approach reduces computational overhead by requiring phase shift calculations only for the primary RIS, while intermediary RISs use pre-designed codewords based on the primary surface selection. We consider two methods for this configuration: a simple linear approach for LoS components and an optimization-based approach for all multipath components. The linear model is less expensive and suitable for LoS-dominated scenarios, reducing computational complexity. The optimization-based approach, though computationally more intensive, optimizes phase shifts for all paths, enhancing spatial multiplexing gains. Comparing these methods helps understand their advantages and limitations in various conditions. For an RIS $j$ focusing on an RIS $i$ ($i, j \in \{1, \dots, I\}$ and $i \neq j$), we derive the phase shift configuration of RIS $j$ to maximize the signal energy received by RIS $i$ from the BS via RIS $j$ using both methods. This process is repeated for all RISs $j$ to derive the codewords for all $i \neq j$. These codewords construct the required codebooks for each RIS, allowing quick and efficient reconfiguration to focus signals towards another RIS.

\subsection{Linear Phase Shifts}

The linear phase shift model maximizes the RIS response function for signals arriving from incidence angles $\mathbf{\Psi}_t^*$ towards reflection angles $\mathbf{\Psi}_r^*$. Revisiting the design in \cite{140}, let $\mathbf{\phi}_i=[e^{j\varphi_{i,1}},\dots,e^{j\varphi_{{i,N_i}}}]^T\in \mathbb{C}^{N_i \times 1}$ be the vector of $N_i$ reflection coefficients of RIS $i$. The phase shift $\varphi_{i,n_{i}}$ of the $(n_{ix},n_{iz})$-th element is:
\begin{align}\label{Lphshift}
    e^{j\varphi_{i,n_{i}}} =& e^{j\varphi_{{i,n_{ix},n_{iz}}}} \nonumber\\=& e^{-j\kappa d_x A_x(\mathbf{\Psi}_t^*, \mathbf{\Psi}_r^*)n_{ix}}e^{-j\kappa d_z A_z(\mathbf{\Psi}_t^*,\mathbf{\Psi}_r^*)n_{iz}}.
\end{align}
The linear response function for a pair $(\mathbf{\Psi}_t,\;\mathbf{\Psi}_r)$ is:
\begin{align} \label{gLi}
    g_{i_L}(\mathbf{\Psi}_t&,\mathbf{\Psi}_r|\mathbf{\Psi}_t^*,\mathbf{\Psi}_r^*)\nonumber\\
    =& \Bar{g}_{uc_i}\sum_{n_{ix}=0}^{N_{ix}-1}\sum_{n_{iz}=0}^{N_{iz}-1} e^{j\kappa d_x(A_x(\mathbf{\Psi}_t, \mathbf{\Psi}_r)-A_x(\mathbf{\Psi}_t^*, \mathbf{\Psi}_r^*))n_{ix}} \nonumber \\
    &\quad \quad \quad \times e^{j\kappa d_z(A_z(\mathbf{\Psi}_t,\mathbf{\Psi}_r)-A_z(\mathbf{\Psi}_t^*, \mathbf{\Psi}_r^*))n_{iz}},
\end{align}
where the maximum value is: $g_{{i_L}_{max}}=g_{i_L}(\mathbf{\Psi}_t^*,\mathbf{\Psi}_r^*|\mathbf{\Psi}_t^*,\mathbf{\Psi}_r^*)=\Bar{g}_{uc_i}N_{ix}N_{iz}=\Bar{g}_{uc_i}N_i$.

To design the inter-RIS focusing codebook using the linear phase shift model, consider an RIS $j$ reflecting signals from the BS towards an RIS $i,\; i\neq j$. With $\mathbf{\Psi}_{r_{BSj}}^{1}$ as the LoS path AoA at RIS $j$ from the BS and $\mathbf{\Psi}_{t_{ji}}^{1}$ as the LoS path AoD from RIS $j$ to RIS $i$, the reflection coefficient of the $(n_{jx},n_{jz})$-th element of RIS $j$ is: 
\begin{align}
    e^{j\varphi_{{j,n_{jx},n_{jz}}}} =& e^{-j\kappa d_x A_x(\mathbf{\Psi}_{r_{BSj}}^{1},\mathbf{\Psi}_{t_{ji}}^{1})n_{jx}}\nonumber\\ & \times e^{-j\kappa d_z A_z(\mathbf{\Psi}_{r_{BSj}}^{1},\mathbf{\Psi}_{t_{ji}}^{1})n_{jz}}.
\end{align}
This phase shift configuration denoted $\mathbf{\phi}_{j \rightarrow i}$ maximizes the RIS response function for the chosen angles, focusing the LoS path from the BS to RIS $j$ towards the RIS $j$-RIS $i$ LoS path. Repeating this for all RISs $i\neq j$, we obtain the linear focusing codebook of RIS $j$:
\begin{equation} \label{CDBKlinear}
    \mathbf{\mathcal{B}}_{j_{Linear}} \triangleq \{\mathbf{\phi}_{j \rightarrow 1},\dots, \mathbf{\phi}_{j \rightarrow j-1},\mathbf{\phi}_{j \rightarrow j+1},\dots, \mathbf{\phi}_{j \rightarrow I}\}. 
\end{equation}
This method provides a low-complexity approach for an inter-RIS focusing codebook, optimal for pure LoS channels. For multipath channels, an optimization-based approach is needed to maximize signal energy from/towards all paths.

\subsection{Optimisation-Based Phase Shifts}

We investigate an optimization-based approach for designing our inter-RIS focusing codebook. The goal is to identify phase shifts that enhance the power of the reflected signals, represented by the squared magnitude of the RIS response function, across all (AoA, AoD) pairs for all paths of the BS-RIS $j$ and RIS $j$-RIS $i$ links. The RIS response function in \eqref{gr} is reformulated as:
\begin{equation}
    g_j(\mathbf{\Psi}_t,\mathbf{\Psi}_r)=\Bar{g}_{uc_j} \mathbf{\phi}_j^H \mathbf{x}_j(\mathbf{\Psi}_t,\mathbf{\Psi}_r) \otimes \mathbf{z}_j(\mathbf{\Psi}_t,\mathbf{\Psi}_r),
\end{equation}
with
\begin{align}
\mathbf{x}_j(\mathbf{\Psi}_t,\mathbf{\Psi}_r) \triangleq & [1,\dots, e^{j\kappa d_xA_x(\mathbf{\Psi}_t, \mathbf{\Psi}_r)(N_{jx}-1)}], \\
\mathbf{z}_j(\mathbf{\Psi}_t,\mathbf{\Psi}_r)\triangleq& [1,\dots, e^{j\kappa d_z A_z(\mathbf{\Psi}_t, \mathbf{\Psi}_r)(N_{jz}-1)}].
\end{align}

We determine a codeword for each RIS $j$ that maximizes the signal energy reflected towards every other RIS $i$. To account for signals across all paths, we aim to maximize the RIS response function for all $L_{Bj}$ incident directions and $L_{ji}$ reflected paths. The optimization problem is modeled as follows:
\begin{equation} \label{focusingOpt}
\begin{aligned}
	&\max_{(\mathbf{\phi}_{j})_1,\dots,(\mathbf{\phi}_{j})_{N_j},\gamma}\quad  \gamma \\
	\textrm{subject to} \quad &  \textrm{C1:}  \; |g_j(\mathbf{\Psi}_{r_{Bj}}^{l_2},\mathbf{\Psi}_{t_{ji}}^{l_1})|^2>\gamma, \; \forall l_1, \; l_2 ,\\
  \quad & \textrm{C2:}\; |(\mathbf{\phi}_{j})_{n_j}|=1,\quad \forall  n_j.
\end{aligned}
\end{equation}
The reflected power is given by:
\begin{equation}
\begin{aligned}
    |g_j&(\mathbf{\Psi}_{r_{Bj}}^{l_2},\mathbf{\Psi}_{t_{ji}}^{l_1})|^2 \\
    &= |\Bar{g}_{uc_j} \mathbf{\phi}_j^H \mathbf{x}_j(\mathbf{\Psi}_{r_{Bj}}^{l_2},\mathbf{\Psi}_{t_{ji}}^{l_1}) \otimes \mathbf{z}_j(\mathbf{\Psi}_{r_{Bj}}^{l_2},\mathbf{\Psi}_{t_{ji}}^{l_1})|^2.
\end{aligned}
\end{equation}
Constraint C2 imposes a unit-modulus constraint on each RIS element, ensuring that the phase shift is applied without any amplification or attenuation. Solving this optimization problem yields two main results. First, we find the maximum value of $\gamma$, ensuring that the response function's square magnitude for all incident and reflected paths exceeds a certain threshold. This enhances even the weakest paths, improving the performance of the double reflected BS-RIS $j$-RIS $i$ link. Second, we identify the optimal phase shifts $\varphi_{{j,n_j}_{j \rightarrow i}}^*,\; \forall n_j$, which correspond to the codeword $\mathbf{\phi}_{j \rightarrow i}^*$ that maximizes $\gamma$. Applying these phase shifts to RIS $j$ increases the signal energy reflected towards RIS $i$, enhancing the desired reflection paths. To obtain the complete codebook of $I-1$ codewords for the $j^{th}$ RIS, we solve optimization problem \eqref{focusingOpt} for all RISs $i\in {1,\dots,j-1,j+1,\dots ,I}$. This results in the inter-RIS signal focusing codebook $\mathbf{\mathcal{B}}_{j_{Opt}}$, defined as:
\begin{equation} \label{CDBKOpt}
\mathbf{\mathcal{B}}_{j_{Opt}} 
    \triangleq \{\mathbf{\phi}_{j \rightarrow 1}^*,\dots, \mathbf{\phi}_{j \rightarrow j-1}^*,\mathbf{\phi}_{j \rightarrow j+1}^*,\dots, \mathbf{\phi}_{j \rightarrow I}^*\}. 
\end{equation}
This procedure is repeated for each RIS to determine all codebooks $\mathbf{\mathcal{B}}_{i_{Opt}}$, for all $i\in \{1,\dots,I\}$. However, the optimization problem in \eqref{focusingOpt} is non-convex, making a globally optimal solution unattainable. To address this, we adopt a sub-optimal approach using Semidefinite Programming relaxation (SDR), similar to the method in \cite{138}. We first reformulate the problem to fit an SDP framework. Despite this, the problem retains a non-convex constraint, which we relax for a tractable solution. The final solution is obtained by solving the relaxed problem and restoring the original non-convex constraint, allowing us to find a locally optimal solution. We transform problem \eqref{focusingOpt} by defining $\mathbf{\Omega}_j = \mathbf{\phi}_j \mathbf{\phi}_j^H\in \mathbb{C}^{N_j \times N_j}$ and using the identity:
\begin{align} \label{eq:identity}
|\mathbf{\phi}_j^H &(\mathbf{x}_j(\mathbf{\Psi}_{r_{Bj}}^{l_2},\mathbf{\Psi}_{t_{ji}}^{l_1}) \otimes \mathbf{z}_j(\mathbf{\Psi}_{r_{Bj}}^{l_2},\mathbf{\Psi}_{t_{ji}}^{l_1}))|^2
\nonumber\\= &tr \bigg( \mathbf{\Omega}_j^H (\mathbf{x}_j(\mathbf{\Psi}_{r_{Bj}}^{l_2},\mathbf{\Psi}_{t_{ji}}^{l_1}) \otimes \mathbf{z}_j(\mathbf{\Psi}_{r_{Bj}}^{l_2},\mathbf{\Psi}_{t_{ji}}^{l_1})) \nonumber \\ &\times(\mathbf{x}_j(\mathbf{\Psi}_{r_{Bj}}^{l_2},\mathbf{\Psi}_{t_{ji}}^{l_1}) \otimes \mathbf{z}_j(\mathbf{\Psi}_{r_{Bj}}^{l_2},\mathbf{\Psi}_{t_{ji}}^{l_1}))^H\bigg). 
\end{align}
Using this identity, we reformulate the optimization problem \eqref{focusingOpt} as:
\begin{equation} \label{focusingOptSDP}
\begin{aligned}
	&\max_{ \mathbf{\Omega}_j,\gamma}\quad  \gamma \\
\quad \quad	\textrm{subject to} \quad &  \textrm{C1:}  \; G_j(\mathbf{\Omega}_j,\mathbf{\Psi}_{r_{Bj}}^{l_2},\mathbf{\Psi}_{t_{ji}}^{l_1})>\gamma, \; \forall l_1, \; \forall l_2 ,\\
  \quad & \textrm{C2:}\; \text{Rank}(\mathbf{\Omega}_j) = 1,\; \textrm{C3:}\; \mathbf{\Omega}_j \succeq 0,\\
\quad & \textrm{C4:} \; \text{Diag}(\mathbf{\Omega}_j) = \mathbf{1}_{N_j},
\end{aligned}
\end{equation}
where
\begin{align}
    G_j(\mathbf{\Omega}_j,\mathbf{\Psi}_{r_{Bj}}^{l_2},\mathbf{\Psi}_{t_{ji}}^{l_1}) =&|\Bar{g}_{uc_j}|^2 tr\bigg( \mathbf{\Omega}_j^H \big(\mathbf{x}_j(\mathbf{\Psi}_{r_{Bj}}^{l_2},\mathbf{\Psi}_{t_{ji}}^{l_1}) \nonumber \\
    &\otimes \mathbf{z}_j(\mathbf{\Psi}_{r_{Bj}}^{l_2},\mathbf{\Psi}_{t_{ji}}^{l_1})\big) \big(\mathbf{x}_j(\mathbf{\Psi}_{r_{Bj}}^{l_2},\mathbf{\Psi}_{t_{ji}}^{l_1}) \nonumber\\ &\otimes \mathbf{z}_j(\mathbf{\Psi}_{r_{Bj}}^{l_2},\mathbf{\Psi}_{t_{ji}}^{l_1})\big)^H\bigg).
\end{align}
Constraints C2 and C3 ensure $\mathbf{\Omega}_j = \mathbf{\phi}_j \mathbf{\phi}_j^H$ after optimization, while C4 guarantees $|(\mathbf{\phi}_{j})_{n_j}|=1$ for all $n_j$. The optimization problem \eqref{focusingOptSDP} remains non-convex due to the rank-one constraint, making a globally optimal solution unattainable. By relaxing C2, we obtain a convex SDP problem that can be solved optimally using tools like CVX \cite{159}. The solution may still be infeasible regarding the non-relaxed problem. Various methods exist to restore the rank-one constraint, with one reliable method being rank reduction via eigenvalue decomposition (EVD) \cite{161}. The procedure is as follows: assuming $\mathbf{\Omega}_j^*$ is the optimal solution of the SDR problem, apply EVD:
\begin{equation}
\mathbf{\Omega}_j^* = \mathbf{U}_j^* \mathbf{\Lambda}_j^* \mathbf{U}_j^{*H},
\end{equation}
where $\mathbf{U}_j^*$ contains eigenvectors and $\mathbf{\Lambda}_j^*$ is a diagonal matrix of eigenvalues. Construct a rank-one approximation using the leading eigenvector:
\begin{equation}
    \mathbf{\Omega}_{j_{rank_1}} = \lambda^*_1 \mathbf{u}^*_1 \mathbf{u}_1^{*H},
\end{equation}
where $ \lambda^*_1 $ is the largest eigenvalue and $ \mathbf{u}^*_1$ is the corresponding eigenvector. The sub-optimal solution for $\mathbf{\phi}_j$ in \eqref{focusingOpt} is given by:
\begin{equation}
    (\mathbf{\phi}_j)_{n_j}=\frac{(\sqrt{\lambda^*_1 }\mathbf{u}^*_1)_{n_j}}{|(\sqrt{\lambda^*_1 }\mathbf{u}^*_1)_{n_j}|},\; \forall n_j\in\{1,\dots,N_j\}. 
\end{equation}

\section{Numerical Results}
\begin{figure}[t]
	    \centering
		\includegraphics[width=42mm]{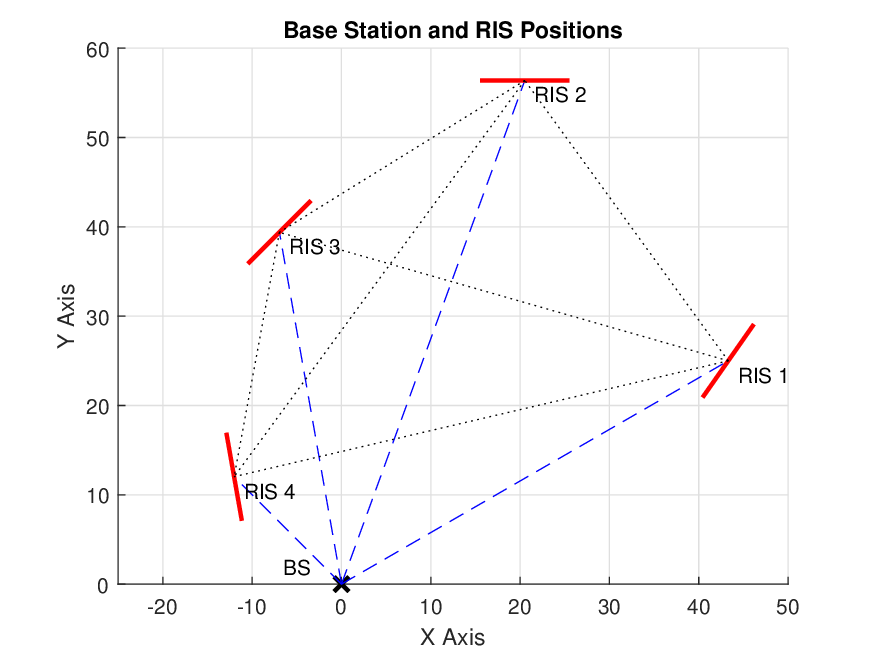}
		\caption{Cooperative RISs Layout.}
		\label{layout}
\end{figure}

To demonstrate the performance of our cooperative inter-RIS focusing codebook, we consider a URPA BS with $M = 10 \times 5$ antenna elements and $I = 4$ URPA RISs. The distances from the BS to the RISs are $\mathbf{d}_{Bi} = [50, 60, 40, 20]$ meters. All units are at the same height, resulting in elevation angles of arrival and departure of $90^\degree$. For the LoS rays, the azimuth AoA and AoD are $\mathbf{\phi}_{t_{BS}}^{1} = [30\degree, 70\degree, 110\degree, 135\degree]$ and $\mathbf{\phi}_{r_{BS}}^{1} = [145\degree, 90\degree, 45\degree, 10\degree]$. The distances $d_{ij}$ between RIS pairs $i$ and $j$ ($i \neq j$), along with the azimuth AoA and AoD for LoS paths $\phi_{t_{ij}}^{1}$, $\phi_{t_{ji}}^{1}$, $\phi_{r_{ij}}^{1}$, and $\phi_{r_{ji}}^{1}$, are calculated using trigonometric relationships involving $d_{Bi}$, $\phi_{t_{BSi}}^{1}$, and $\phi_{r_{BSi}}^{1}$. For other rays indexed by $l_1 \in \{2, \ldots, L_{Bi}\}$ and $l_2 \in \{2, \ldots, L_{ij}\}$, $\forall i,\;j,\; i\neq j$, azimuth angles are randomly generated from a uniform distribution centered around the LoS angles within a range $\Delta_a$. Both $L_{Bi}$ and $L_{ij}$ are set to 3 for all $\forall i,\;j,\; i\neq j$. The inter-element distances for the BS antennas and RIS elements are $d_x = d_z = \frac{\lambda_c}{4}$, with rectangular unit cell dimensions $L_{xi} = L_{zi} = \frac{\lambda_c}{4}$ for all $i$. Fig. \ref{layout} illustrates the layout.

\begin{figure}[t]
    \centering
    \begin{subfigure}[t]{0.49\linewidth} 
        \centering
        \includegraphics[width=1.1\linewidth]{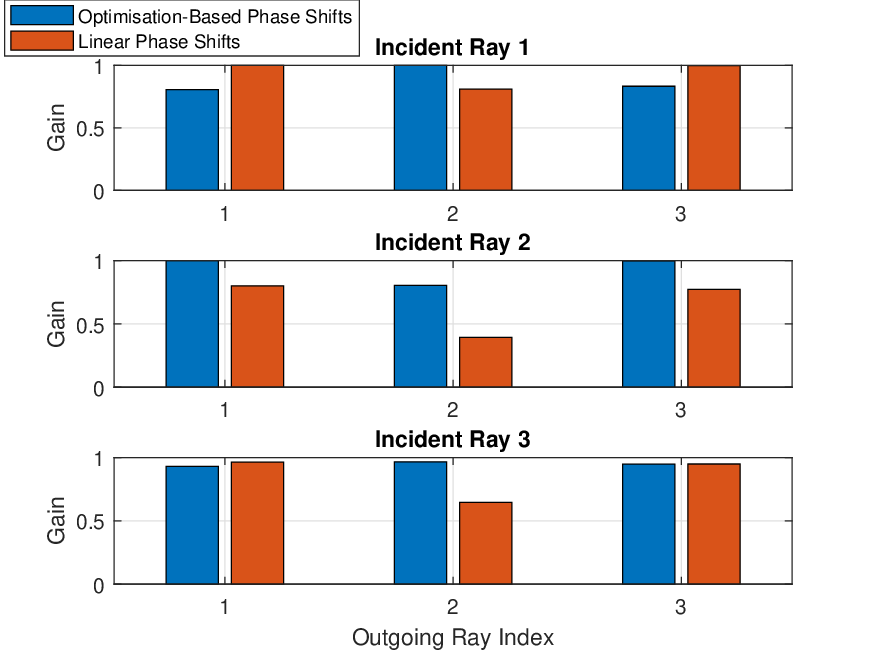}
        \caption{$N_i=7\times7$, $\Delta_a=10\degree$.}
        \label{fig:sub1}
    \end{subfigure}
    \begin{subfigure}[t]{0.49\linewidth} 
        \centering
        \includegraphics[width=1.1\linewidth]{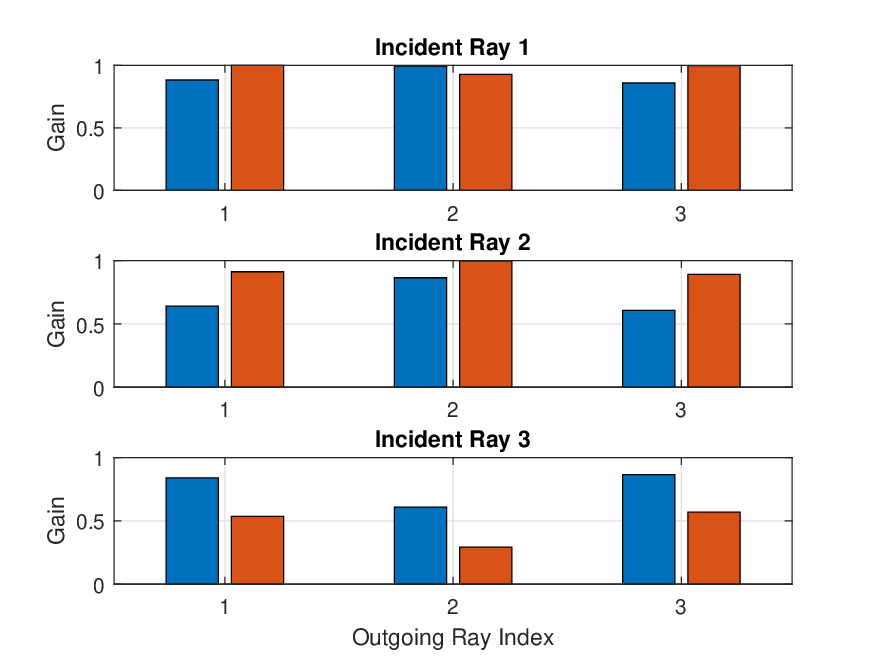}
        \caption{$N_i=7 \times 7$, $\Delta_a=20\degree$.}
        \label{fig:sub2}
    \end{subfigure}

    \begin{subfigure}[t]{0.49\linewidth} 
        \centering
        \includegraphics[width=1.1\linewidth]{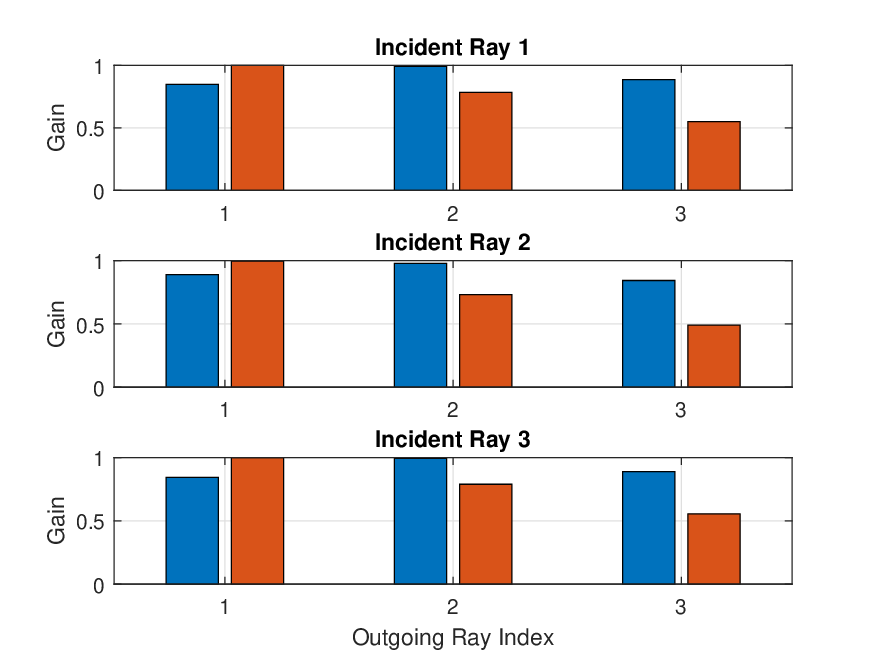}
        \caption{$N_i=10 \times 10$, $\Delta_a=10\degree$.}
        \label{fig:sub3}
    \end{subfigure}
    \begin{subfigure}[t]{0.49\linewidth} 
        \centering
        \includegraphics[width=1.1\linewidth]{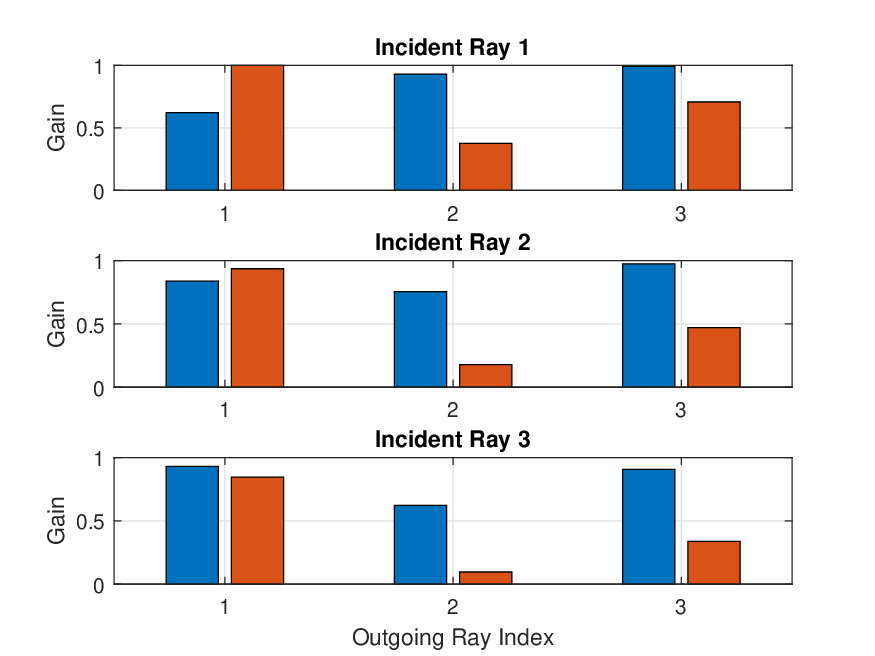}
        \caption{$N_i=10 \times 10$, $\Delta_a=20\degree$.}
        \label{fig:sub4}
    \end{subfigure}

    \caption{Normalised RIS response function gain for $L_{Bi}=L_{ij}=3$ with RIS $i=1$ focusing towards RIS $j=3$.}
    \label{fig:mainfigure}
\end{figure}

The analysis focuses on the gain of the RIS response function, quantified by the square of the normalized gain, $\mid\frac{g_i}{\Bar{g}_{i}}\mid^2$, where $\Bar{g}_{i}=\Bar{g}_{uc_i}N_i$ represents the maximum amplitude of the RIS response function. The primary aim is to assess the RIS's ability to channel signals from the BS to intended directions with minimal leakage for efficient and targeted signal steering. The study examines the interaction of $L_{Bi}$ BS signal rays reflected off RIS $1$ towards $L_{ij}$ rays connected to RIS $3$. Additionally, it explores interactions with other RIS configurations (RIS $2$ and $4$) to evaluate how effectively the codebook achieves signal directionality without interference.

In Fig. \ref{fig:mainfigure}, gains are depicted for angle spreads $\Delta_a=10\degree$ and $20\degree$, evaluated using RIS configurations with $N_i=7 \times 7$ and $10 \times 10$ reflecting elements. The analysis demonstrates that linear phase shifts consistently provide maximum gain for LoS to LoS ray paths. This outcome aligns with the design intention, confirming the efficacy of linear phase shifts in scenarios where direct paths are dominant. For $\Delta_a=10\degree$ and $N_i=7\times 7$ (Fig. \ref{fig:sub1}): Both the linear and optimization-based designs exhibit significant gains across the range of arrival and departure angles. This performance showcases their effectiveness when the angle spread is narrow, ensuring most rays fall within the optimal radiation lobe. For $\Delta_a = 20^\circ$ and $N_i=7 \times 7$ (Fig. \ref{fig:sub2}): As the angle spread increases, the linear codebook's efficacy diminishes, particularly for NLoS rays. In contrast, the optimization-based codebook maintains well-distributed gains. This robust performance is due to its strategy of maximizing the minimum gain across all angle pairs, thus offering superior adaptability to wider angle variations. Furthermore, increasing the number of RIS elements enhances directivity, narrowing the main beam lobe. This effect can be seen in Figs. \ref{fig:sub3} and \ref{fig:sub4}, where a larger RIS configuration ($N_i = 10 \times 10$) is employed. While this boosts directivity, it results in poorer gain distribution with linear designs due to beam narrowing. The optimization-based approach, however, adapts well to these changes, maintaining balanced gain distribution even with higher directivity.
\begin{figure}[t]
    \centering
    \begin{subfigure}[t]{0.49\linewidth} 
        \centering
        \includegraphics[width=1.1\linewidth]{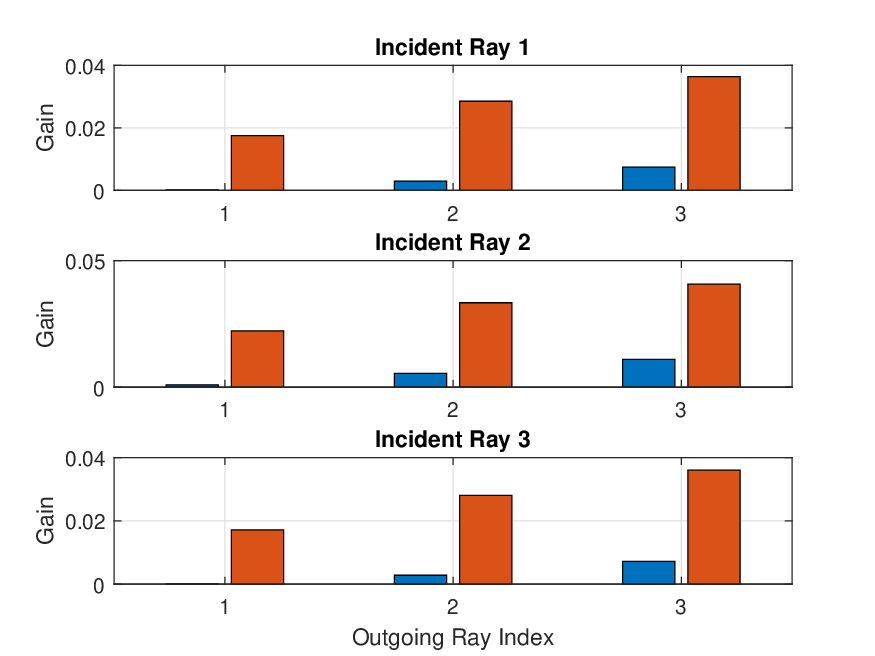}
        \caption{Leakage towards RIS $2$.}
        \label{leakage2}
    \end{subfigure}
    \begin{subfigure}[t]{0.49\linewidth}
        \centering
        \includegraphics[width=1.1\linewidth]{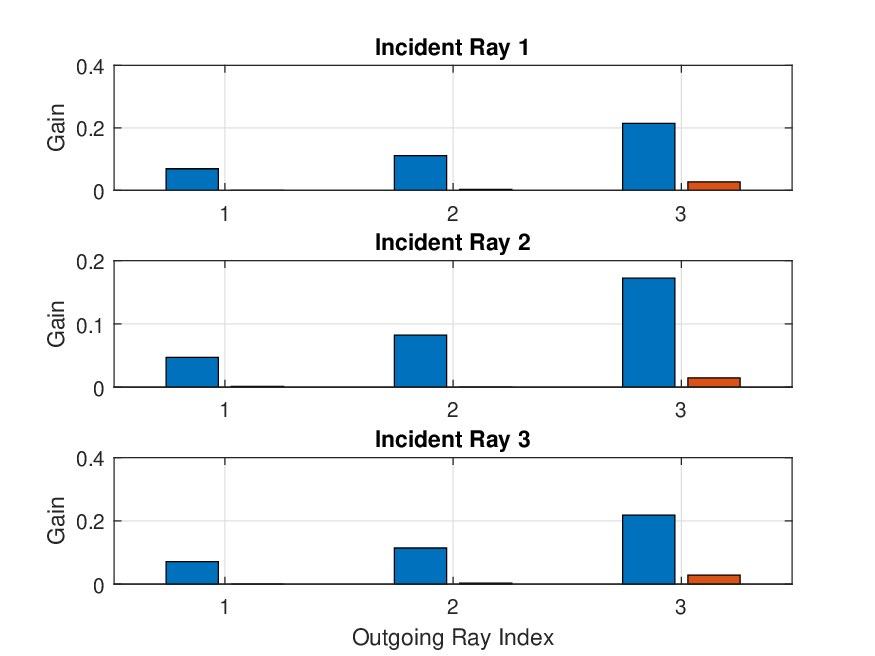}
        \caption{Leakage towards RIS $4$.}
        \label{leakage4}
    \end{subfigure}

    \caption{Normalized gain towards undesired paths with RIS $1$ focusing on RIS $2$, $N_i=10\times 10$, $\Delta_a=10\degree$.}
    \label{interference}
\end{figure}

The study further evaluates gains towards RISs $2$ and $4$ under the phase shift configurations designed for RIS $ 1$, focusing on RIS $3$. RIS $1$ to RIS $2$: As shown in Fig. \ref{leakage2}, the gain towards rays connecting RIS $1$ and $2$ remains below $4\%$ for both codebook designs. This low leakage highlights the scheme's efficacy in steering signals accurately, minimizing undesired spillover between RISs. RIS $1$ to RIS $4$: Fig. \ref{leakage4} reveals up to a $20\%$ leakage with the optimization-based codebook for the third ray between RISs $1$ and $4$. This increase is due to the proximity of RISs $3$ and $4$, leading to unseparated paths between RIS $1-3$ and RIS $1-4$. This issue can potentially be mitigated by deploying larger RISs with enhanced directivity and optimizing the positional and angular arrangement of the RISs.

\begin{figure}[t]
    \centering
    \begin{subfigure}[t]{0.49\linewidth} 
        \centering
        \includegraphics[width=1.1\linewidth]{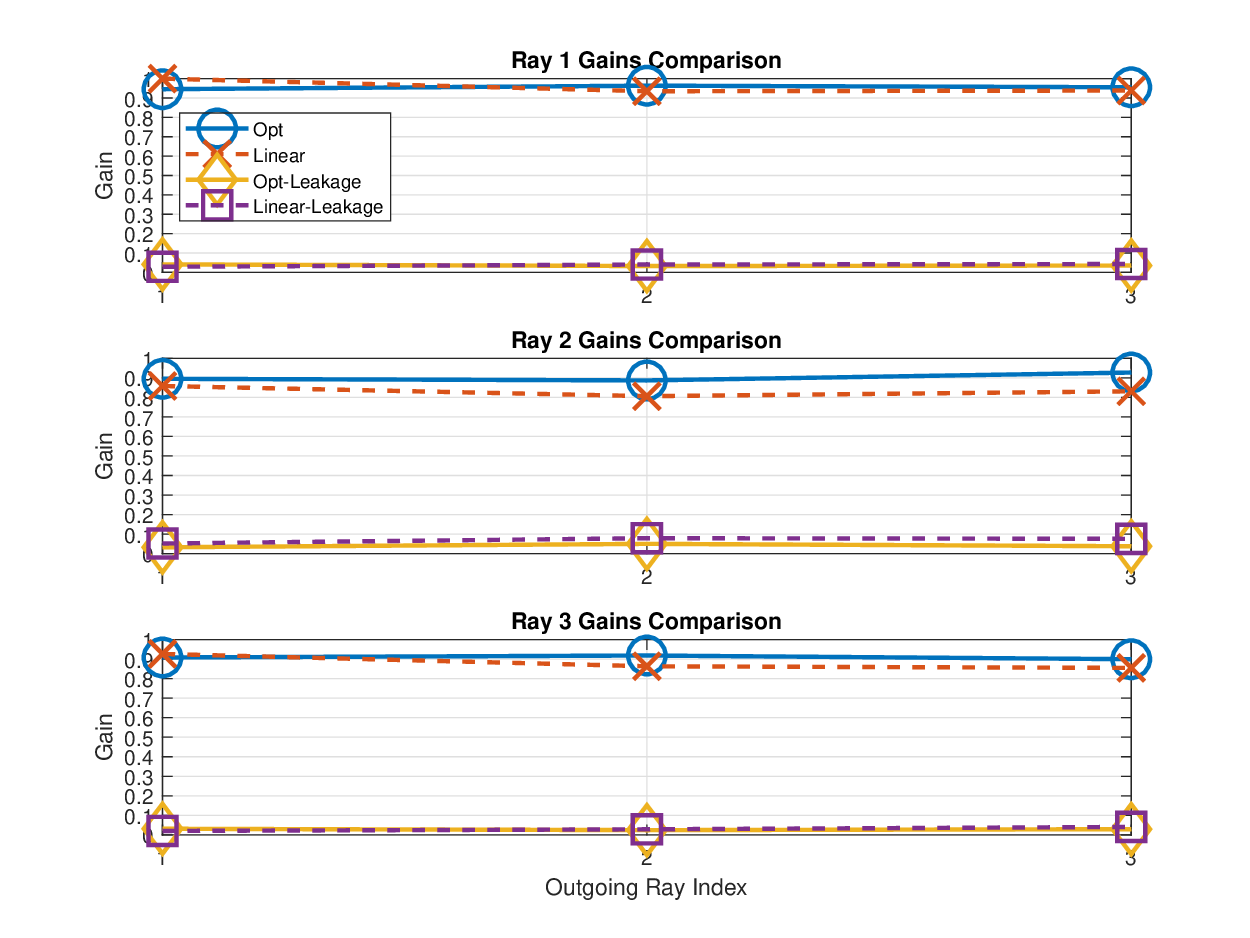}
        \caption{$N_i=7\times 7$, $\Delta_a=10\degree$.}
        \label{fig:sub11}
    \end{subfigure}
    \begin{subfigure}[t]{0.49\linewidth}
        \centering
        \includegraphics[width=1.1\linewidth]{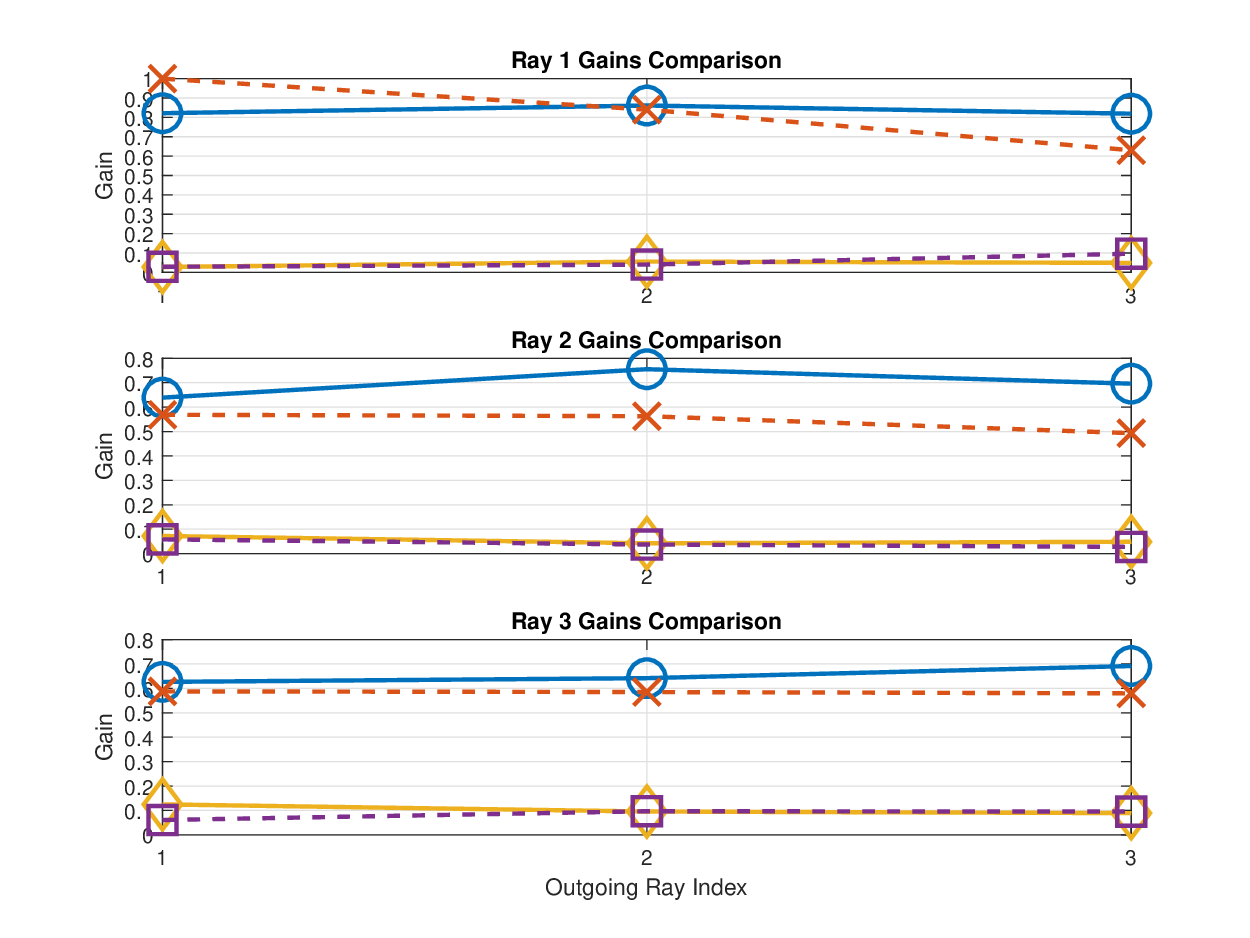}
        \caption{$N_i=7\times 7$, $\Delta_a=20\degree$.}
        \label{fig:sub22}
    \end{subfigure}

    \begin{subfigure}[t]{0.49\linewidth} 
        \centering
        \includegraphics[width=1.1\linewidth]{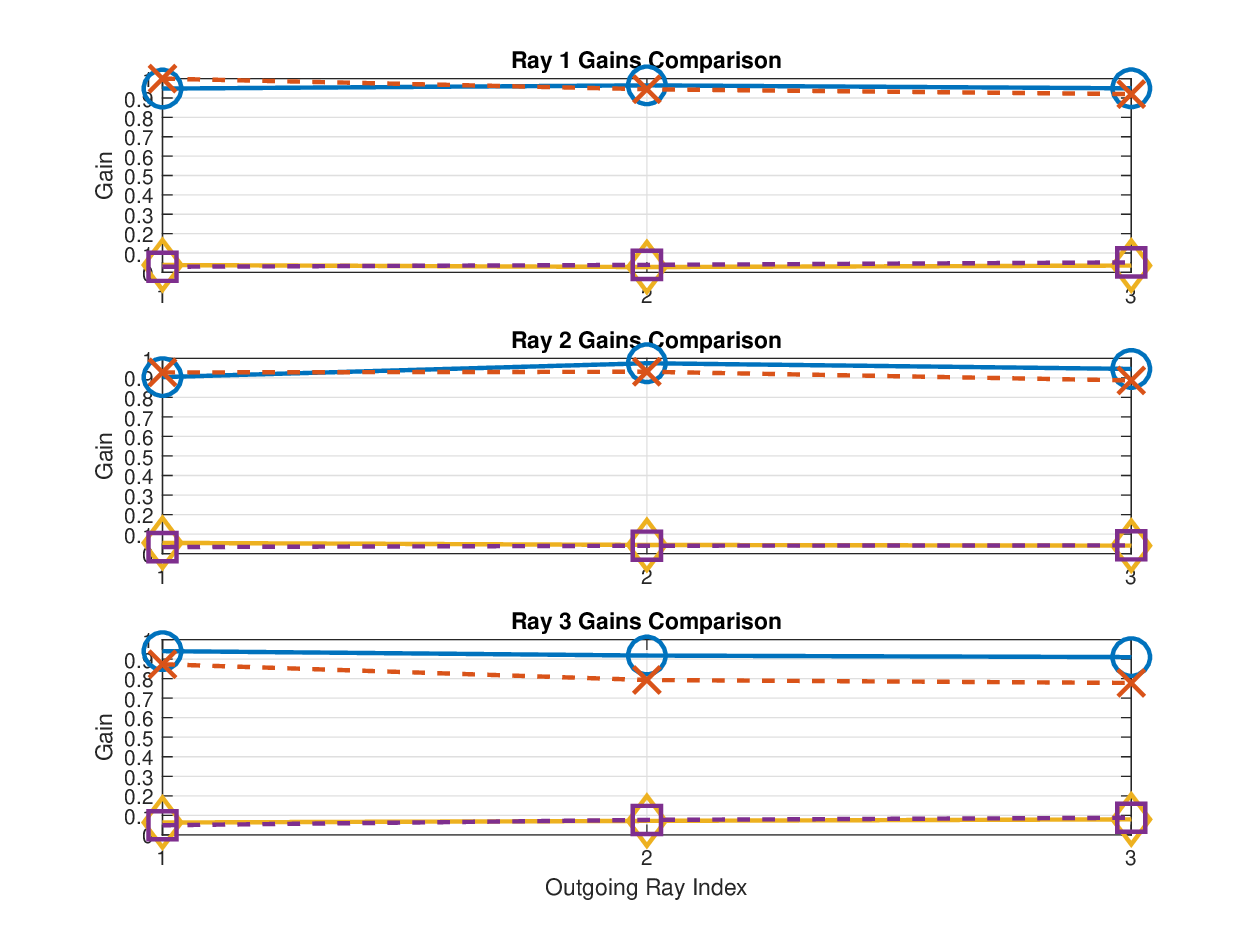}
        \caption{$N_i=10\times 10$, $\Delta_a=10\degree$.}
        \label{fig:sub33}
    \end{subfigure}
    \begin{subfigure}[t]{0.49\linewidth} 
        \centering
        \includegraphics[width=1.1\linewidth]{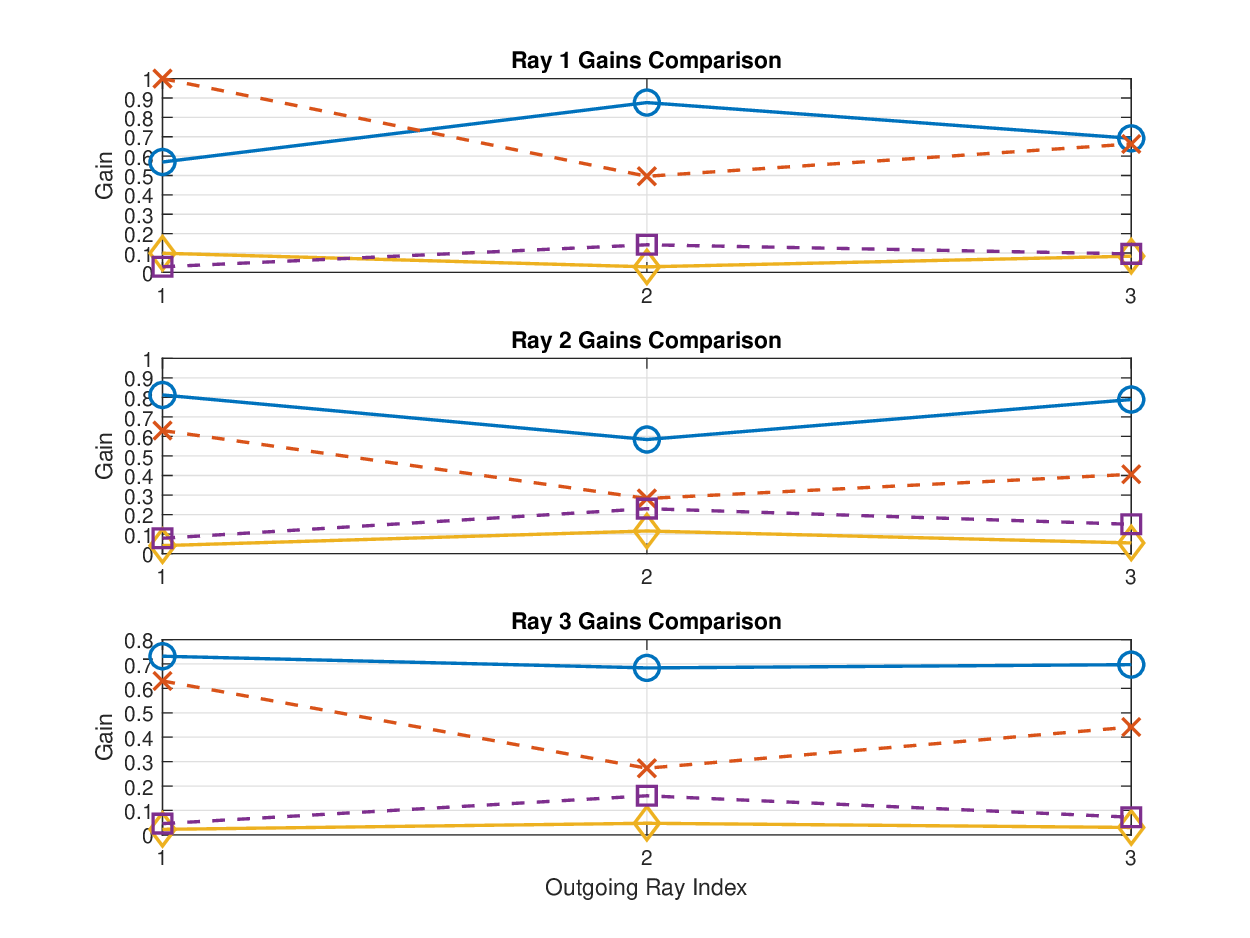}
        \caption{$N_i=10\times 10$, $\Delta_a=20\degree$.}
        \label{fig:sub44}
    \end{subfigure}

    \caption{Average normalized gain for $L_{Bi}=L_{ij}=3$.}
    \label{avgains}
\end{figure}

The comprehensive effectiveness of the codebook designs is evaluated in Fig. \ref{avgains}, which plots the average gain across all RIS pairs. The analysis includes averaging gains for intended paths and leakage across unintended RIS pairings, thereby providing a holistic view of codebook efficacy in the cooperative RIS system. Both linear and optimization-based codebooks perform well under smaller angle spreads and RIS sizes, but the optimization-based design proves superior at handling larger $\Delta_a$ and $N_i$. This versatility affirms the proposed optimization-based approach's reliability when dealing with varied operational conditions, ultimately enhancing the spatial multiplexing capabilities of the cooperative system by ensuring robust exploitation of available transmission paths. Furthermore, the average leakage remains remarkably low across all scenarios considered, with the optimization-based codebook maintaining minimal leakage even as $\Delta_a$ increases.

\section{Conclusion}
In this letter, we introduced the design of an inter-RIS focusing codebook aimed at enhancing the performance of multi-RIS aided MU-MIMO systems. By employing both linear and optimization-based phase shifts, our approach effectively maximizes signal energy reflection among RISs. The numerical results indicated that the optimization-based codebook outperforms the linear design in multipath scenarios, while the linear design provides the optimal configuration in pure LoS conditions.

\bibliography{References} 
\bibliographystyle{IEEEtran}

\end{document}